\documentclass[twocolumn,showpacs]{revtex4}

\usepackage{dcolumn}
\usepackage{amsmath}
\usepackage{graphicx}

\begin{document}

\title{Jamming transition of a granular pile below the angle of repose}

\author{S. Deboeuf$^1$, E.M. Bertin$^2$, E. Lajeunesse$^1$, O. Dauchot$^2$} 
\affiliation{$~^1$ IPGP, CNRS/INSU UMR 7580, 75005 Paris, France}
\affiliation{$~^2$ DSM/SPEC, CEA Saclay, 91 191 Gif sur Yvette, France}

\date{\today}

\begin{abstract}
We study experimentally the relaxation towards mechanical equilibrium of a granular pile which has just experienced an avalanche and discuss it in the more general context of the granular jamming transition. Two coexisting dynamics are observed in the surface layer: a short time exponential decay consisting in rapid and independent moves of grains and intermittent bursts consisting in spatially correlated moves lasting for longer time. The competition of both dynamics results in long-lived intermittent transients, the total duration of which can late more than a thousand of seconds. We measure a two-time relaxation function, and relate it via a simple statistical model to a more usual two-time correlation function which exhibits strong similarities with auto-correlation functions found in aging systems.  Localized perturbation experiments also allow us to test the pile surface layer receptivity.
\end{abstract}


\maketitle

\section{Introduction}

Dry, non-cohesive granular materials can flow like a liquid but can also sustain under the influence of gravity a finite angle of repose $\theta_r$. Given the large number of degrees of freedom involved in the dynamics, it is tempting to think in terms of phase transition: for a pile slope $\theta>\theta_r$, surface layers flow downhill, whereas for $\theta<\theta_r$, these layers \lq\lq freeze\rq\rq and the pile behaves like a solid. As a matter of fact, Bak et al.~\cite{Bak87} proposed the idea that a granular pile \lq\lq self-organizes\rq\rq into a critical state at the angle of repose. However, a number of experimental works~\cite{Evesque88,Jaeger89,Liu91,Grasselli97,Courrech03} have studied surface instability of a granular heap, namely avalanches, and have led to the conclusion that such avalanches do not behave in a critical manner at all.

In recent years, it has also been suggested that the jamming transition of granular materials could be the analog of a glass transition~\cite{Jaeger92,Liu98}. Further works on mean field glass models have  developed this analogy and tried to unify concepts that had emerged in both fields, such as the dynamical temperature defined from fluctuation dissipation relations, and that related to Edwards' statistical ensemble~\cite{Edwards94,Kurchan00,Barrat00,Berthier01,Makse01}.

In this paper, we reconsider the avalanching pile problem but, instead of focusing on avalanches statistics, we study the relaxation of the pile towards mechanical equilibrium following an avalanche. We first show that the dynamics exhibits non-trivial relaxations and is much slower than expected given the \lq\lq microscopic\rq\rq time-scale. We observe intermittent bursts which temporarily reactivate the pile activity and identify them with correlated movements in space and time. Perturbation experiments allow us to confirm the emergence of strong spatial correlations when increasing the pile slope. Finally we extract in the unperturbed case a two-time relaxation function, which can be shown to rescale in an aging like manner, including a `reparametrization' of time .
suggested by a simple model that we introduce in order to describe our experimental data. Moreover, this model allows to relate the above relaxation function to a two-time correlation function which shares strong similarities with auto-correlation functions appearing in aging systems.

The experimental setup is described in section II whereas the observations are reported in section III. Section IV is then devoted to the presentation of the model. Finally, the experimental results are discussed in section V in the light of the model, emphasizing possible connections with glassy dynamics.

\section{Experimental setup}

\begin{figure}[htb]
\centering\includegraphics[width=8cm]{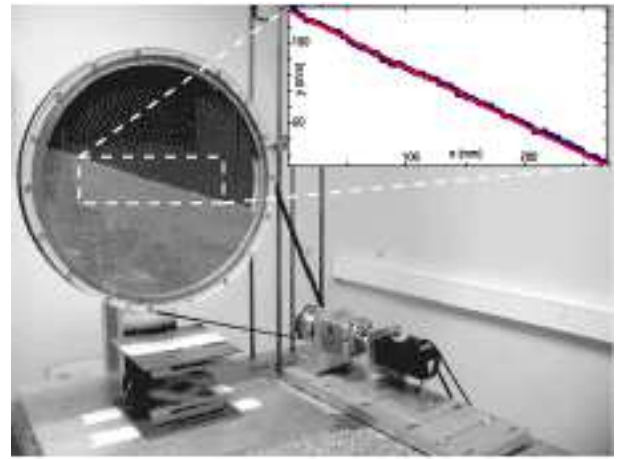}
\caption{Picture of the experimental setup. The inset displays a typical acquisition of the pile slope.}
\label{setup}
\end{figure}

The experimental setup shown on figure~\ref{setup} consists in a rotating drum of internal diameter $D= 450 {\rm mm}$ and thickness $\delta=22 {\rm mm}$, half-filled with steel beads of diameter $d=3\pm 0.025 {\rm mm}$ and density $\rho=7.8 {\rm g.cm^{-3}}$. 

\noindent
The friction coefficient and coefficient of restitution of the beads are respectively $\mu=0.2$ and $e=0.92$. A DC motor drives the drum clockwise or anticlockwise at angular velocities as low as $\Omega = 0.01^{\circ}/s$. One can fix the pile slope to any given value smaller than the repose angle by simply braking the drum.
The pile is lightened by a continuous halogen lamp located far enough from the drum to prevent heating. Also the lab temperature is regulated at a constant value $20^{\circ}C$. The experimental area of interest is filmed either by a standard CCD camera ($768 \times 572$ pixels at $25 {\rm i/s}$) or by a fast one ($480 \times 210$ pixels at $1000 {\rm i/s}$) aligned along the axis of the drum. In both cases, the camera focus is set to have a spatial resolution of $1 {\rm mm}$ per pixel. Image acquisition and processing allows us to extract the pile slope, the surface roughness and the pixels where a displacement as small as $25 \mu{\rm m}$ as occurred.  

The surface detection takes advantage of the gradient of the light intensity fluctuations which is maximum at the surface of the pile $S(x)$. The averaged slope of the pile $\theta$ is then deduced from a linear least square fitting of $S(x)$ while the surface roughness $r$ is given by the average deviation around the mean slope. Beads displacements are detected using an image difference method already described in~\cite{Bonamy03}. In principle measuring the intensity difference between two consecutive images of the pile allows to identify the region where beads displacements have occurred. In practice, one needs to reduce the noise caused by lighting inhomogeneities, camera intrinsic vibrations and acquisition noise. This is achieved by acquiring two consecutive sequences of $n$ images: $n=25$ (respectively $n=200$) for the standard (respectively the fast) camera. The two sequences are averaged resulting in two frames. Substracting them and binarizing them so as to remove any remaining noise leads to the detection of the area where displacements have occurred. $n$ is chosen according to a compromise between the noise/signal ratio and the temporal resolution. The sensitivity of this detection method is calibrated by translating the camera with a micrometric translator: any displacement larger than $20 \mu{\rm m}$ is spotted unambiguously. The method takes advantage of the whole range of the acquisition dynamics to detect displacements much smaller than the pixel size. The counterpart is that only the existence of such small displacement is detected, but not its amplitude.

\section{Experimental results}

\subsection{Repose and avalanche angles}

We first measure the repose and avalanche angles $\theta_r$ and $\theta_a$ of the granular pile. The drum is rotated at a very low velocity ($\Omega = 0.01^{\circ}/{\rm s}$). The pile slope then follows an intermittent regime of macroscopic avalanches \cite{Rajchenbach90, Caponeri95, Courrech03}. The pile slope $\theta(t)$ and the surface roughness $r(t)$ are extracted every $5 {\rm s}$ from two experimental runs of $12$ hours each.

\begin{figure}[htb]
\centering\includegraphics[width=8cm]{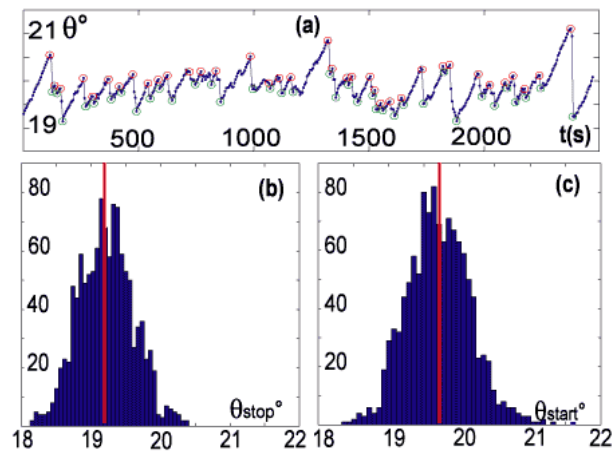}
\caption{(a): pile slope angle $\theta(t)$ $(^{\circ})$ as a function of time.
(b), resp.(c): histograms of $\theta_{stop}$, resp. $\theta_{start}$.}
\label{pente}
\end{figure}

\noindent
A typical sequence of the pile slope evolution $\theta(t)$ is displayed on figure \ref{pente}(a). In such a low rotation regime, $\theta(t)$ increases linearly with time at the rate $\Omega$ until an avalanche occurs at $\theta_{start}$ and the slope relaxes toward a stopping angle $\theta_{stop}$. The pile roughness remains on the order of half a grain diameter and no correlations between $r(t)$ and $\theta(t)$ or their temporal derivatives emerge. Filtering the events for which the pile slope variation $(\theta_{start} - \theta_{stop})< 0.1^{\circ}$ corresponds to a local slope rearrangements implying only a few beads, we ended with a total number of a thousand of events identified as avalanches. The resulting histograms of $\theta_{start}$  and  $\theta_{stop}$ -- figure \ref{pente}(b),(c) -- exhibit exponential tails, which could indicate non-trivial fluctuations in $\theta_{start}$ and $\theta_{stop}$. Still, in absence of a better definition, we choose to estimate the repose and avalanche angles $\theta_r$ and $\theta_a$ as the averaged values of respectively $\theta_{stop}$ and $\theta_{start}$ leading to $\theta_r=19.2^\circ \pm 0.2^\circ$ and $\theta_a=19.7^\circ \pm 0.2^\circ$.  The relatively small difference between $\theta_a$ and $\theta_r$ can be explained by the large restitution coefficient of our steel beads.

\subsection{Relaxation dynamics}

In order to investigate the relaxation process of the granular pile following an avalanche, the drum is rotated at constant angular velocity ($1^{\circ}/s$) during a few minutes so that a number of successive avalanches have occurred. The drum is then stopped just after a given avalanche ($\theta=\theta_{stop}$) and rapidly rotated backwards down to the chosen working angle $\theta < \theta_{stop}$, where it is braked firmly. The relaxation process is recorded with the standard CCD camera.
Consecutive sequences of $n=25$ images of the pile are taken every $15 {\rm s}$. Beads displacements occurring in between two sequences are detected following the procedure described in section II. As a result one obtains a time series of binary images where black pixels indicate positions where displacements have occurred. 

\begin{figure}[htb]
\centering\includegraphics[width=8cm]{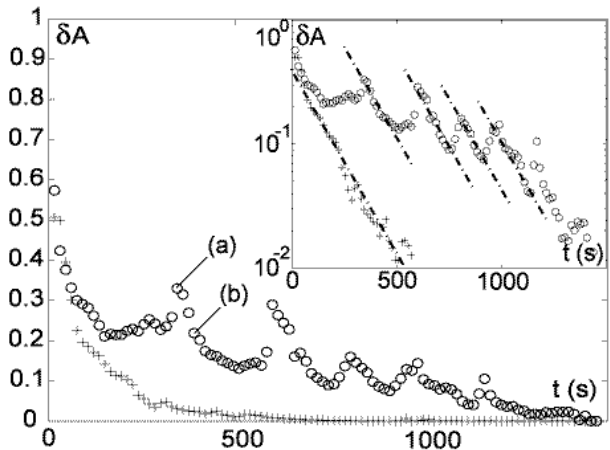}
\caption{The mobile volume fraction evolution $\delta A(t)$ for two different realizations at similar pile slope : ($\circ$) $\theta=15^{\circ}$; ($+$) $\theta=16.5^{\circ}$. Inset is the log-lin plot of the same data. Notice the exponential decay rate, which is identical in the monotonous case ($+$) and in the intermittent case ($\circ$).}
\label{relax1}
\end{figure}
\begin{figure}[htb]
\centering\includegraphics[width=8cm]{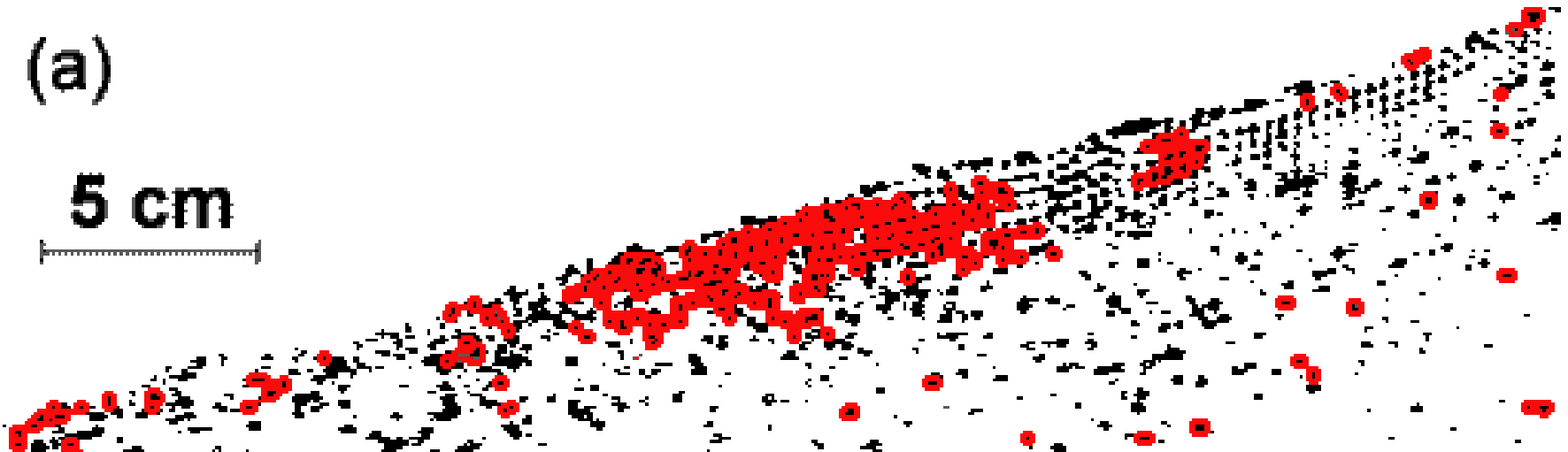}
\centering\includegraphics[width=8cm]{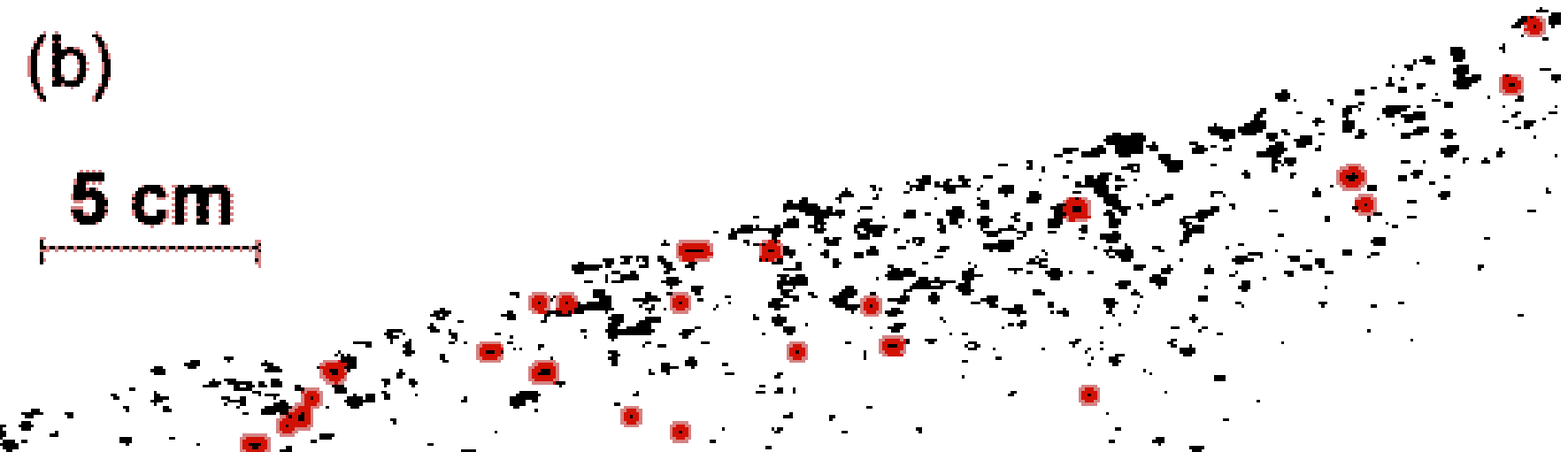}
\caption{Displacements in the pile at time step (a) and (b) labelled on figure~\ref{relax1}. The dark pixels corresponds to positions where a displacement has occurred in the $15 {\rm s}$ preceeding the given time step. The grey overlay indicates the pixels among those, where displacements have occurred successively during $30 {\rm s}$ following the given time step (see text for details).}
\label{relax2}
\end{figure}

\noindent
The \lq\lq mobile volume fraction\rq\rq of the pile $\delta A(t)$ (i.e. the fraction of beads that have moved between the two acquisitions) is computed as the total number of black pixels normalized with the total number of pixels covering the pile. 
Figure~\ref{relax1} displays two typical records of $\delta A(t)$ as observed for two realizations of the pile relaxation with similar slopes $\theta=15^{\circ}$ and $\theta=16.5^{\circ}$. During the very first time steps the relaxation process is identical in both records: the bulk of the pile relaxes rapidly from bottom to top on timescales of the order of the $15 {\rm s}$ delay between two consecutive images acquisitions. It involves isolated beads displacement on time-scale even shorter and of amplitude much smaller than the bead diameter. The relaxation process then slows down in a subsurface layer of thickness of the order of $[10-20]$ beads diameters. In contrast with the bulk, the subsurface layer relaxation may differ significantly from one realization to another as it is the case for the two realizations presented here. In the following, we will focus on the relaxation of the subsurface layer.  In one case, one observes a simple exponential decay of the subsurface layer activity with a characteristic timescale $\tau_{\downarrow}$ which in the present case is of the order of $200{\rm s}$. In the other case, the pile activity evolution is more complex.  One notices intermittent bursts which interrupt periods of exponential decay, with the same time-scale as in first case (see inset of fig~\ref{relax1}). The competition between the exponential relaxation and the reactivation bursts results in an intermittent process occurring on much longer time-scales lasting up to $30$ minutes. 

A visual inspection reveals that the reactivation bursts correspond to collective motion of grain clusters whereas the exponential decay observed in between two bursts or when there is no burst at all involves individual beads displacements. This is illustrated on figure~\ref{relax2}, which displays the inverse binary pictures of two time steps  --(a) during a burst event, (b) during an exponential decay period. The mobile volume fraction illustrated by the relative number of dark pixels, $\delta A$ is of the same order and their spatial distributions are very similar. On these pictures, the gray overlay indicates the areas, where the same pixels have successively record displacements during $45 {\rm s}$ ($15 {\rm s}$ before and $30 {\rm s}$ after the given time step). In the case of the burst event, and in contrast with the exponential decay case, it shows very clearly, that those displacements which last in time are also correlated in space, in the form of a cluster of grains.\\
 
\begin{figure}[htb]
\centering\includegraphics[width=8cm, height=5.0cm]{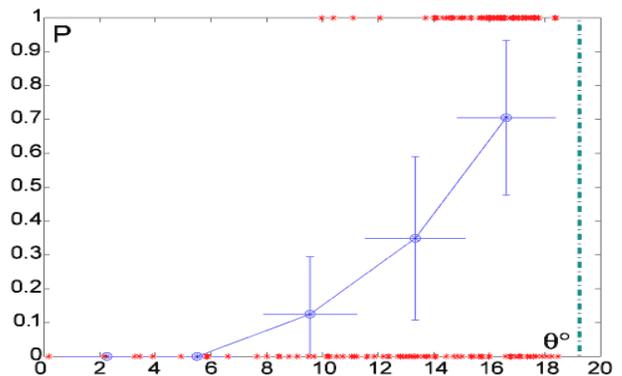}
\caption{Probability of observing an intermittent relaxation dynamics as a function of the pile slope. The crosses show the realizations which have been performed and are set to 0 for simple exponential decay and to 1 for intermittent dynamics. The vertical dash-dotted line indicates the angle of repose $\theta_r=19.2^\circ$.}
\label{intproba}
\end{figure}

We now investigate the influence of the pile slope on the relaxation dynamics by performing $185$ realizations with $\theta_o \in [0^{\circ}, \theta_r]$. As shown on figure~\ref{intproba}, the probability of observing intermittent dynamics increases with the pile slope. No intermittent dynamics could be observed for $\theta_o < 5^{\circ}$ -- most likely a statistical effect.

\begin{figure}[htb]
\hspace{-3mm}\includegraphics[width=8.2cm]{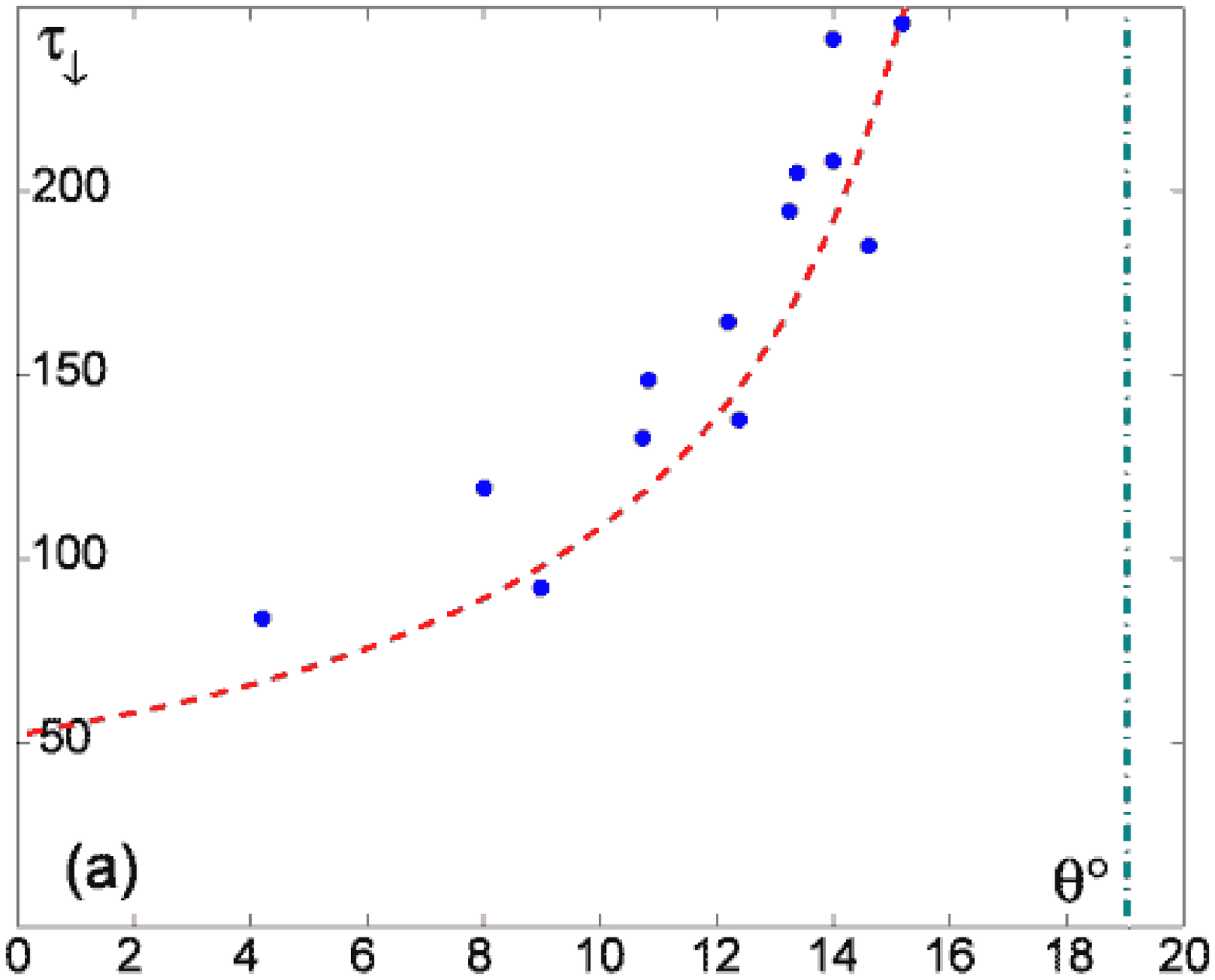}
\vspace{5mm}
\includegraphics[width=8cm]{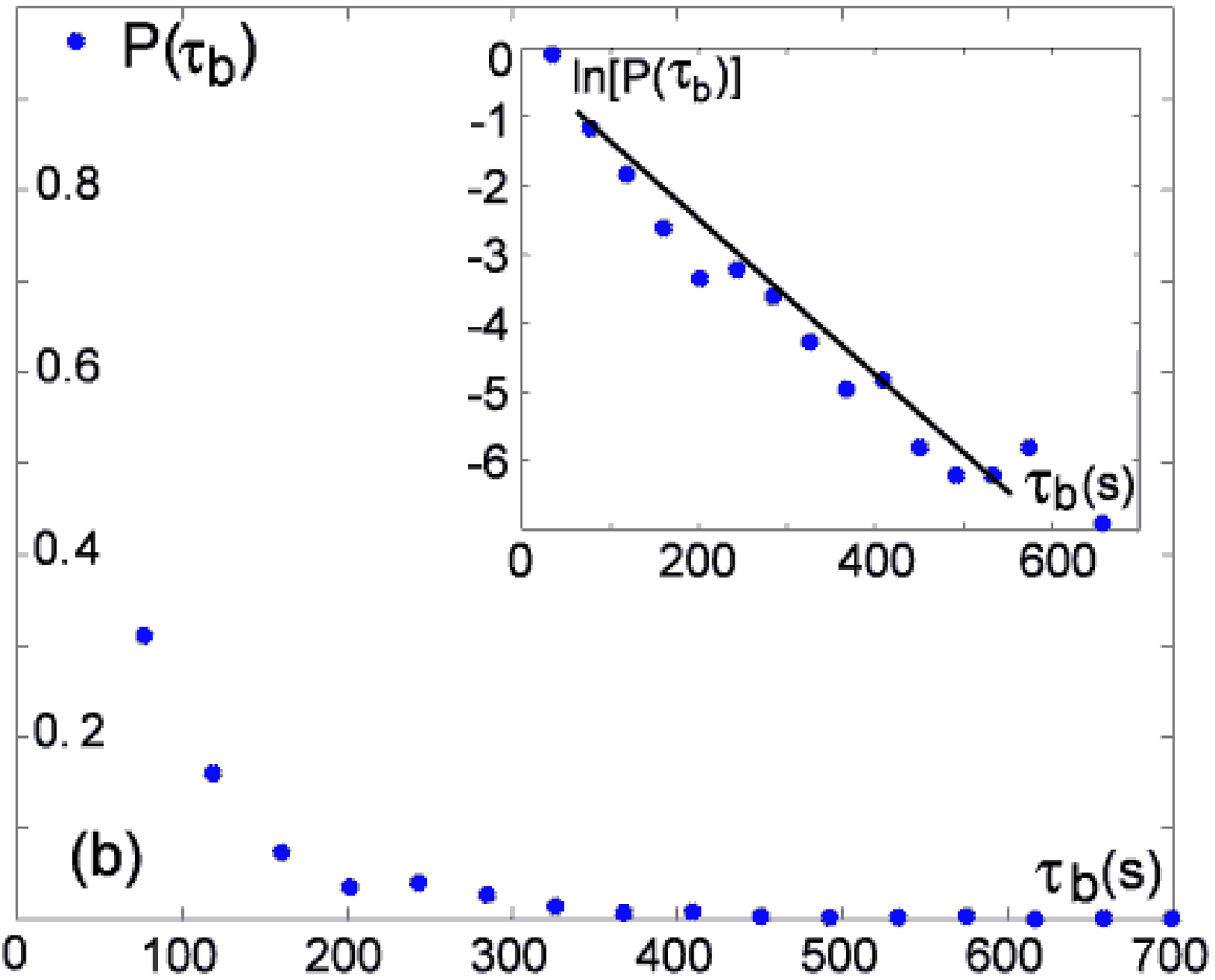}
\caption{(a) Averaged relaxation time for exponential decay realizations as a function of the pile slope and  (b): Probability distribution of the time intervals between bursts in the case of intermittent dynamics.}
\label{times}
\end{figure}

Figure~\ref{times}(a) shows the averaged characteristic relaxation time of the exponential decay $\tau_\downarrow$ as a function of the pile slope. It increases significantly from $50 {\rm s}$ for $\theta=0^{\circ}$ to $250 {\rm s}$ when $\theta$ approaches the repose angle and can be fitted by a power-law in $(\theta_r-\theta)^{-1}$. By contrast, the reactivation process time-scales do not exhibit significant variations with $\theta$. The distribution of time intervals between two successive bursts, shown on figure~\ref{times}(b), exhibits an exponential tail, $P(T)\sim e^{-T/\tau_b}$, where $\tau_b=100 {\rm s}$ is the typical waiting time between two successive bursts. Interestingly one can notice that the pile slope for which the monotonous relaxation time $\tau_\downarrow$  becomes of the order of  the  characteristic time interval between two bursts $\tau_b$ coincides with an angle $\theta\approx 10^\circ$ for which  the probability of observing bursts during the relaxation becomes significant ($>0.1$).  This indicates that bursts only occur when the relaxation process is still active in agreement with direct observations. 

\begin{figure}[htb]
\includegraphics[width=8.2cm]{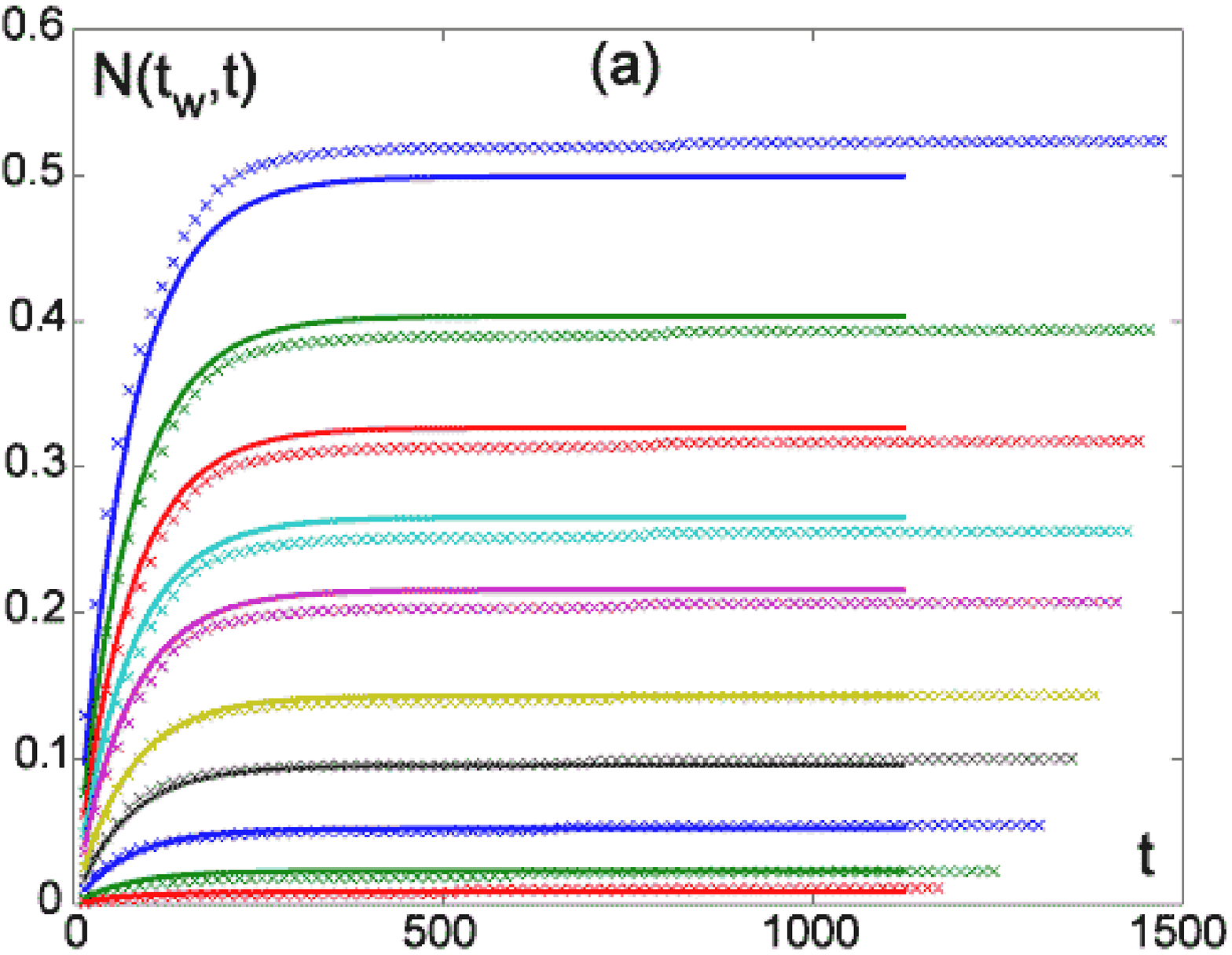}
\vspace{5mm}
\hspace{2mm}\includegraphics[width=8cm]{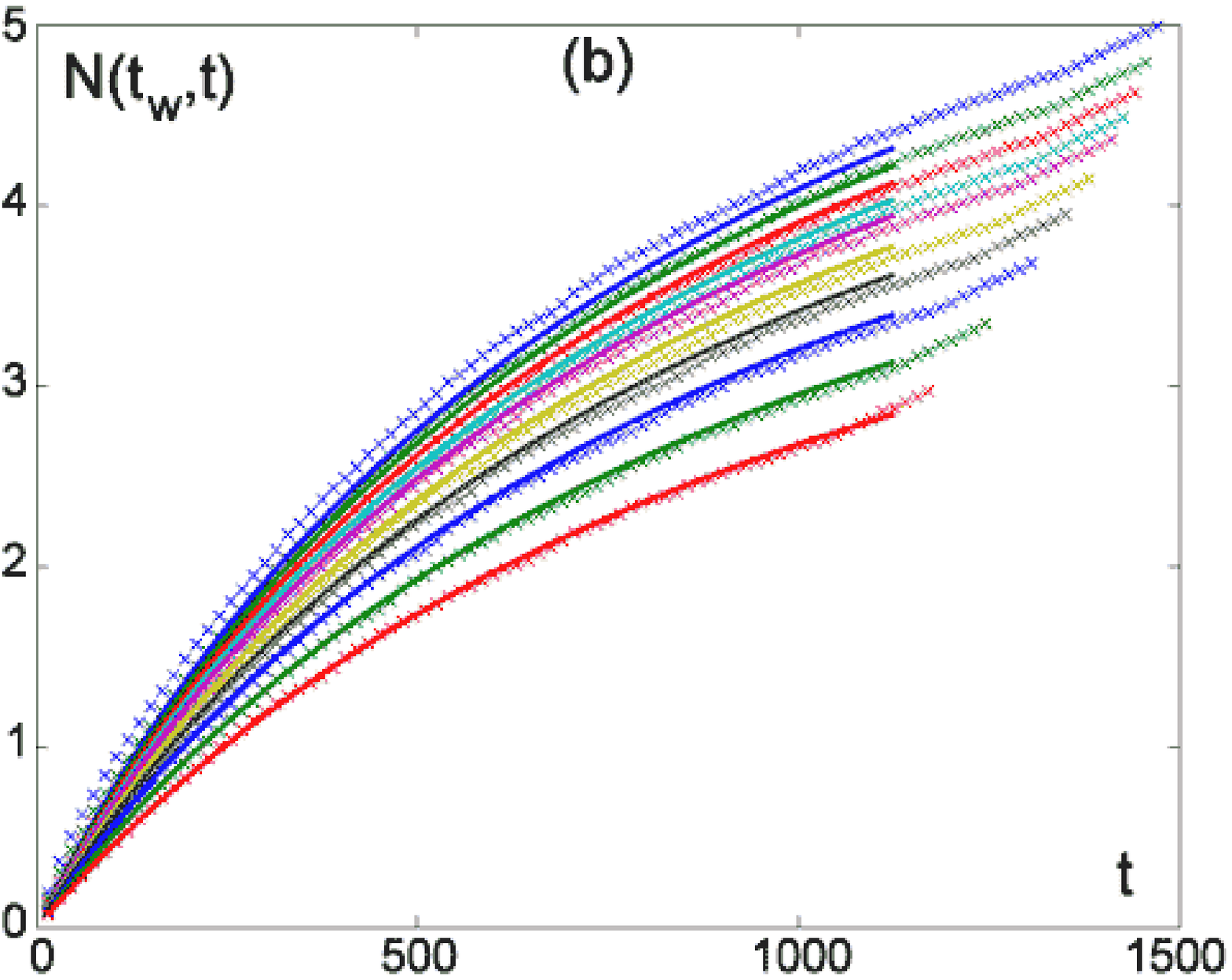}
\caption{$N(t_w,t)$, averaged number of displacements per pixel between times $t_w$ and $t_w+t$ plotted vs.~$t$ increasing $t_w$ (top to bottom): (a) exponential decay case and $\theta<5^\circ$, (b) intermittent dynamics case and $\theta>13^\circ$. $\times$:~experimental data; continuous lines: 3 parameter fit for each set of curves (see text, section IV for details).}
\label{Nexp}
\end{figure}

The relaxation being a non stationnary process, it is described more accurately by measuring a two-time relaxation function. It is convenient experimentally to measure $N(t_w,t)$ the averaged number of displacements per pixel between times $t_w$ and $t_w+t$, given by:

\begin{equation}
N(t_w,t) = \left\langle \int_{t_w}^{t_w+t} \delta A(u) du \right\rangle
\end{equation}

\noindent
where the brackets denote ensemble average over realizations with characteristic time-scales of the same order. We compute $N(t_w,t)$ for both types of relaxation. For the exponential decay case (figure~\ref{Nexp}(a)), we average over all realizations obtained for $\theta<5^\circ$. For the intermittent dynamics case (figure~\ref{Nexp}(b)), we average over the intermittent realizations obtained for $\theta>13^\circ$.  

\noindent
In the first case, $N(t_w,t)$ rapidly saturates whatever the value of $t_w$ and remains smaller than one, which indicates that displacements occur in average less than once at any given position. This is in agreement with the picture of isolated and independent moves as described above for the exponential decay dynamics. In the second case, no saturation occurs. $N(t_w,t)$ increases continuously and becomes much larger than one, which reflects the possibility of several displacements at a given position. In neither cases, could we find an obvious rescaling of time. Figure~\ref{Nexp} also displays fitting curves of our experimental data. We will come back on these fits and their interpretation in section IV.

\subsection{Response of the pile to a localized perturbation}

In order to gather some indications on the state of the subsurface layer during the relaxation, we probe  its response to a localized and instantaneous disturbance. During the pile relaxation, we let a bead fall on the pile surface from a height of $2{\rm cm}$, with zero initial velocity. The whole procedure is recorded by the $1000{\rm i/s}$ camera during $1 {\rm s}$ covering the impact of the incident bead. The image difference method described in section II is applied with two sequences of images acquired just before and after the impact in order to detect the displacements induced by the impact. We fixed the binarization threshold such that the very short delay of the measure ensures not to take into account displacements coming from the relaxation dynamics. Several realizations are performed for each pile slope and the reverse binary images obtained for each realization are averaged in order to obtain a gray-scale picture of the local probability of displacement. 

\begin{figure}[htb]
\centering\includegraphics[width=8cm]{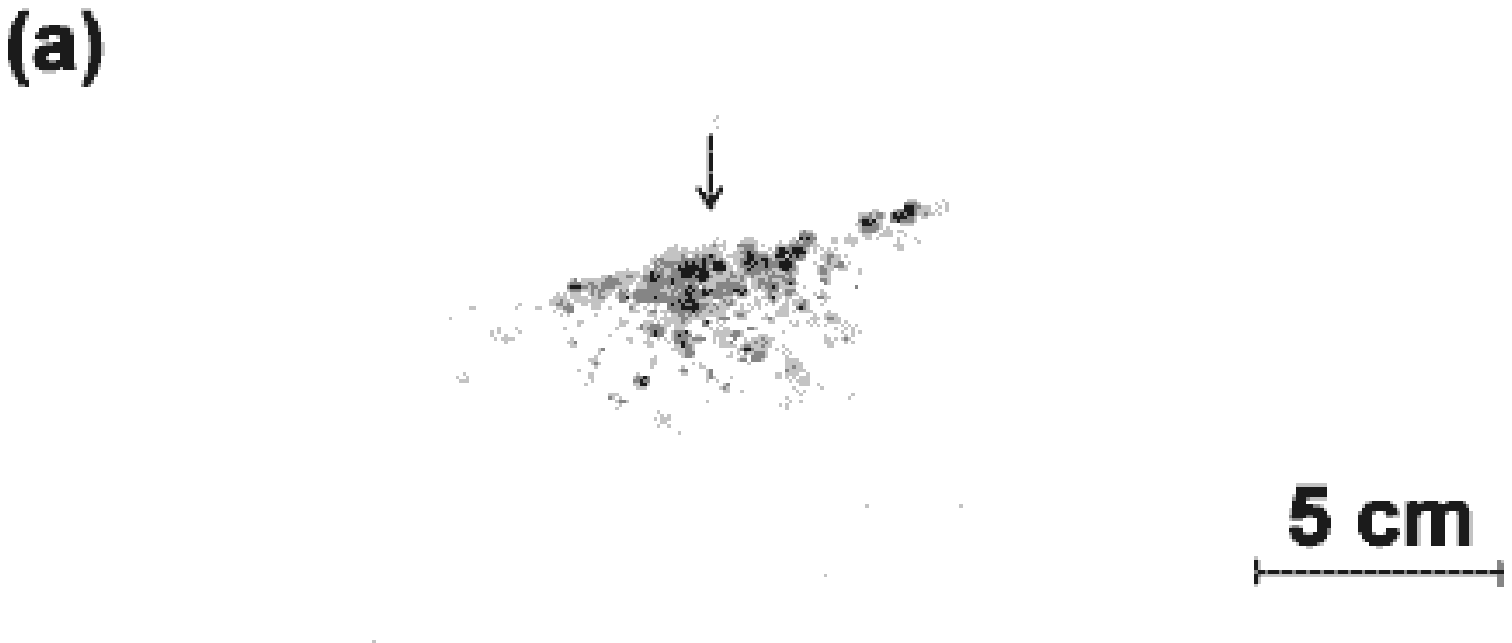}
\centering\includegraphics[width=8cm]{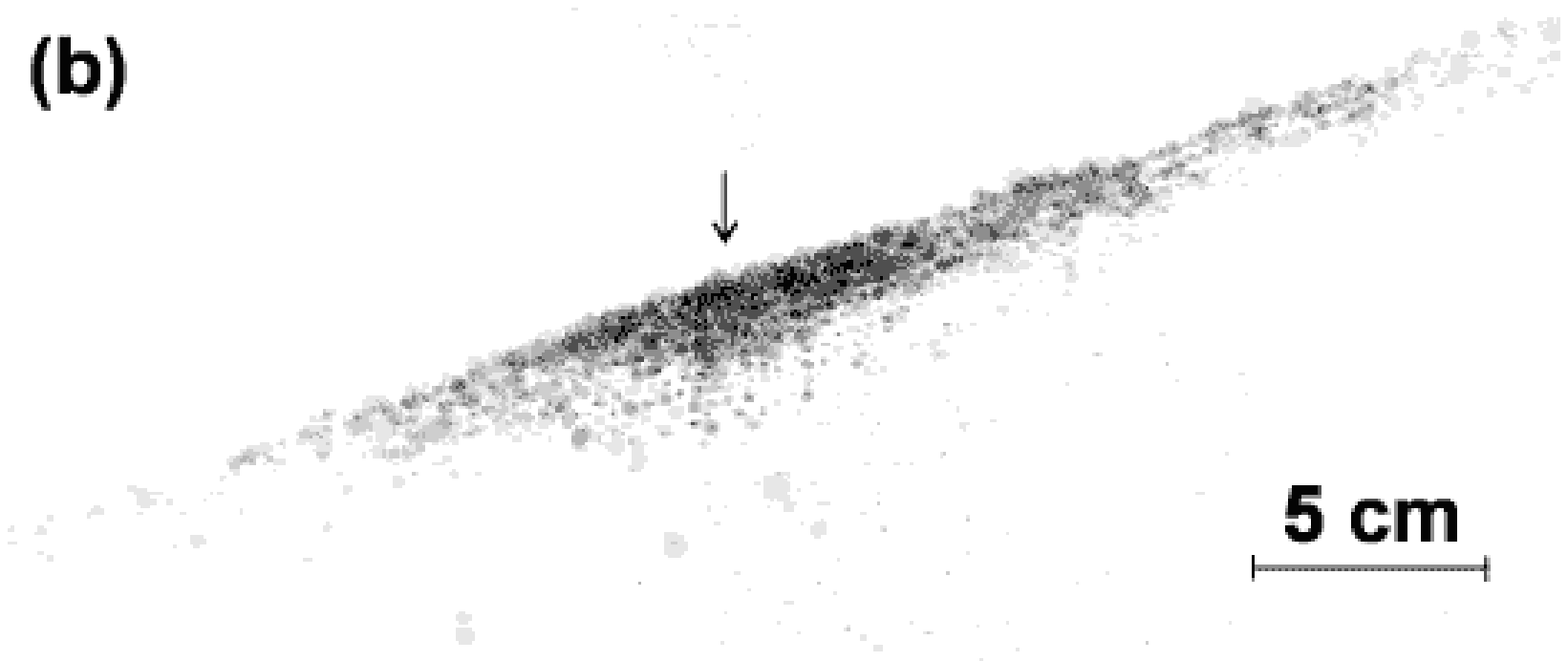}
\centering\includegraphics[width=8cm]{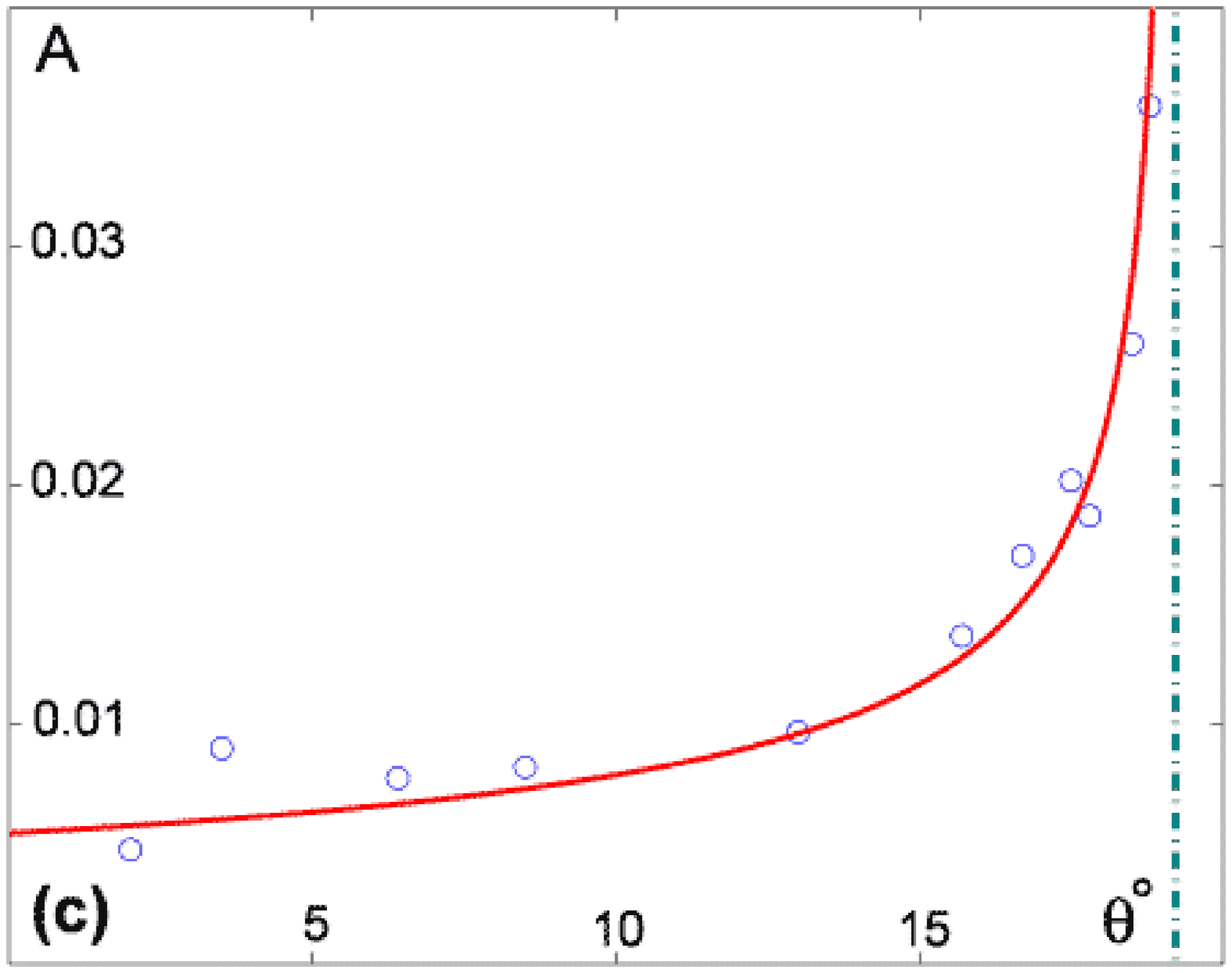}
\caption{Response in displacements of the pile to a bead impact. Grayscale pictures of the local probability of having a displacement for pile slopes to (a) $\theta = 13.2^\circ$ and (b) $\theta = 18.8^\circ$; (the arrow indicates the localization of the bead impact. (c): Impacted area $A(\theta)$ as a function of the slope angle.}
\label{perturbation}
\end{figure}

Two typical responses obtained for pile slopes respectively equal to $\theta = 13.2^\circ$ and $\theta = 18.8^\circ$  are shown on figure~\ref{perturbation}(a) and (b). Note that the depth of penetration of the impacted area is essentially constant of the order of [10-20] beads diameters whatever the slope angle, the same thickness as the subsurface layer where the slowing down of the relaxation dynamics occurs. In contrast, the lateral extension strongly increases with $\theta$ and is larger uphill than downhill resulting in a asymmetrical response. The amplitude of the response is estimated by measuring the area of the impacted zone $A(\theta)$ as a function of the pile slope. As shown on figure~\ref{perturbation}(c) the response  increases strongly with $\theta$ and can be fitted by a power-law in $(\theta_r-\theta)^{-0.5}$. 

\subsection{Summary}

We now summarize all experimental results. Just after the occurrence of an avalanche the granular pile that we study relaxes very rapidly -- less than $15 {\rm s}$ -- in its bulk, but exhibits a much slower relaxation in a subsurface layer of thickness $[10-20]$ beads diameter. In this subsurface layer, exponential decay and reactivation bursts compete resulting eventually in long-lived -- more than $10^3 {\rm s}$ -- intermittent transients depending on the occurrence of bursts during the relaxation process. At the bead scale, the exponential decay is composed of fast and small independent displacements, whereas the reactivation bursts are linked with correlated moves of beads clusters. The exponential decay time $\tau_\downarrow$ increases from $50 {\rm s}$ to $250 {\rm s}$ like $(\theta_r-\theta)^{-1}$. The typical time interval separating two bursts $\tau_b \approx 100 {\rm s}$. Accordingly, the probability of having at least one burst becomes significant when $\tau_\downarrow$  becomes on the order of $\tau_b$, that is for $\theta\approx 10^\circ$. $N(t_w,t)$, the averaged number of displacements per pixel between times $t_w$ and $t_w+t$ behaves very differently depending on the dynamics which is considered -- with or without reactivation events. Finally the instantaneous response of the pile to a given localized disturbance increases sharply in terms of spatial extension when the pile slope approaches $\theta_r$. These observations raise at least two questions of different nature:

\begin{itemize}
\item How can one observe well defined time scales $\tau_\downarrow$ and $\tau_b$ of the same order $10^2 {\rm s}$, given that the only characteristic time that can be defined at the bead scale is $\sqrt{d/g} \approx 10^{-2}{\rm s}$?
\item Is there a simple model which allows a direct calculation of $N(t_w,t)$ and proposes an interpretation of the relaxation dynamics of the pile? Could one find moreover a time rescaling for $N(t_w, t)$, as it often happens in glassy dynamics?
\end{itemize}

Answering the first question, which has to be considered in the context of micro-mechanics, is out of reach of the present experimental setup. Still, recent numerical studies~\cite{Staron02, Radjaipriv} reveal the existence of two types of constitutive structures, namely the strong forces chains network and critical contacts clusters. Whereas the characteristic size of the former remains constant, the size of the latter increases with $\theta$. 
These results suggest an intuitively appealing, although rather speculative parallel. 
A constant $\tau_b$ could be a fingerprint of the underlying role played by the strong force network. Conversely, the increase with $\theta$ of the monotonous relaxation time $\tau_\downarrow$ might be associated with the growth of the critical clusters. Further numerical studies of these collective mechanisms are under progress. Also one should take into account the possible effect of aging at the contact scale.

To deal with the second question, we introduce a simple model in which particles -- to be thought of as beads -- can move under some specific conditions, such that the number of moves $N(t_w,t)$ can be computed analytically. 

\section{Model}

It has been argued above that the measured displacements were due to individual beads relaxation, which were from time to time reactivated by some collective rearrangements. In the following model, we further simplify this physical picture by considering that the duration of the collective rearrangement can be neglected, which amounts to say that the `activity level' $\delta A$ is instantaneously raised when a reactivation burst occurs. To implement this picture, we assume that particles can be in two distinct kinds of states, namely active states labeled from $1$ to $n$ and an inactive state labeled $0$. In a active state, a particle can \lq\lq move\rq\rq spontaneously to any other state, with a rate (probability per unit time) $\alpha$ for transitions to the inactive state $0$ and $\alpha'$ for transitions to the other active states, resulting in a global rate of transition starting from an active state $\gamma = \alpha + (n-1) \alpha'$. These transitions are to be associated to physical displacements (or individual relaxations), and their average number is thus the analog of the experimentally measured quantity $N(t_w,t)$. On the contrary, particles in the inactive state cannot evolve spontaneously, and thus do not contribute to $N(t_w,t)$. In order to account for the collective reactivation process, we introduce an external mechanism which chooses particles at random -- and independently of their state -- and sets them in an active state, with a probability $\nu$ per unit time and per particle. If the chosen particle is already in a active state, this reactivation process has no effect. At this stage, the reactivation mechanism is considered not to involve any displacement of the bead, but rather a collective rearrangement of its environment. For this reason, reactivations are assumed {\it not} to contribute to the average number of moves $N(t_w,t)$.

Introducing the probability $P_i(t)$ to be in state $i$ at time $t$ ($i=0 \ldots n$), the fraction $P_m(t)$ of active particles is given by $P_m(t) = \sum_{i=1}^n P_i(t)$. From this, we deduce the following evolution equation:

\begin{equation}
\frac{dP_m}{dt} =  - \alpha P_m + \nu P_0 = - \alpha P_m + \nu \left(1-P_m \right)
\label{equ1}
\end{equation}
where the first term accounts for the total probability to relax towards the inactive state $0$, and the second one corresponds to the reactivations starting from the inactive state. Let us now introduce the key ingredient of the model: thermal noise being irrelevant for granular media, it is reasonable to think that the reactivation process is driven by some mechanical noise generated by the relaxation of beads in the vicinity of the particle considered (the range of `interaction' might however be quite large, due to the presence of strong correlations, induced for instance by the force chains network). This effect can be taken into account by imposing that the rate $\nu$ of reinjection depends, in a mean-field spirit, on the global number of active particles: $\nu = \nu(P_m)$. We consider here only the simplest functional dependence: $\nu = \mu P_m$, where $\mu$ is a constant. Solving Eq.~(\ref{equ1}) using this assumption leads to:

\begin{equation}
P_m(t) =  \frac{1}{\mu} \frac{d\ln g}{d\, t}
\label{equ2}
\end{equation}

\noindent
where an auxiliary function $g(t)$ has been introduced:

\begin{equation}
g(t) =  (\alpha -\mu) + \mu P_m(0) \left[1 - e^{-(\alpha - \mu)t} \right]
\label{equ3}
\end{equation}

\noindent
From these results, one can compute the average number $N(t_w,t)$ of particle displacements (in the sense defined above) between times $t_w$ and $t_w+t$:

\begin{equation}
N(t_w,t) = \int_{t_w}^{t_w+t} \gamma P_m(t) dt  = \frac{\gamma}{\mu} \ln\left(\frac{g(t_w+t)}{g(t_w)}\right)
\label{equ4}
\end{equation}

\noindent
The exact calculation of $N(t_w,t)$ will allow us to fit our experimental data. In order to relate our results to the more general context of glassy dynamics, it is of interest to calculate in this model the correlation function $C(t_w,t)$ defined as the probability not to have changed state between $t_w$ and $t_w+t$. This correlation decomposes into the sum of the probability not to move starting from an active state at $t_w$ and the probability not to be reactivated given that the particle was inactive at $t_w$. This leads to the following expression for $C(t_w,t)$:

\begin{equation}
C(t_w,t) = P_m(t_w) e^{-\gamma t} + [1-P_m(t_w)] \frac{g(t_w)}{g(t_w+t)} 
\label{correlation}
\end{equation}


We now check the validity of the model by fitting the experimentally measured $N(t_w,t)$ with the expression given in Eq.~({\ref{equ4}).
Whereas the above expressions depend on four parameters, $\alpha$, $\gamma$, $\mu$, the elementary transition rates and $P_m(0)$, the initial fraction of active particles, $N(t_w,t)$ actually depends on three parameters only $\gamma_P=\gamma P_m(0)$, $\delta=\alpha-\mu$ and $\mu_P=\mu P_m(0)$. Extracting these parameters from the experimental data by a unique least-square fit of the full 2-variable measurement of $N(t_w,t)$ for each type of relaxation (i.e. monotonous and intermittent) leads us to the fit displayed on figures~\ref{Nexp}(a),(b) and the experimental estimation of the parameters written on the left side of table~\ref{paramtab}. Figure~\ref{Nscaled} explicitly demonstrates the quality of the rescaling obtained for $N(t_w,t)$ as a function of $g(t_w+t)/g(t_w)$. 

\begin{table}[htb]
\begin{center}
\begin{tabular}{r|r|r|cr|r|r|}
 		&  (a)   & (b) 	&\hspace{2cm} 	&		&   (a)	&  (b)		\\
$\gamma_P$	& 0.0096 & 0.0087	& 		& $1/\alpha$	&  56 s	&  294 s	\\
$\delta$ 	& 0.0134 & 0.0008 	& 		& $1/\delta$	&  75 s	&  1250 s	\\
$\mu_P$ 	& 0.0023 & 0.0014	& 		& $\eta$	&   1	&  4.7		\\
\end{tabular}
\caption{Parameters values extracted from the experimental data for each type of relaxation: (a) the exponential decay and (b) the intermittent dynamics.
Left: direct estimation of the parameters by a unique least-square fit of $N(t_w,t)$.
Right: relaxation time scales $1/\alpha$ and $1/\delta$, and average number $\eta$ of moves per beads (see text for details).}
\label{paramtab}
\end{center}
\end{table}

\begin{figure}[htb]
\includegraphics[width=8cm]{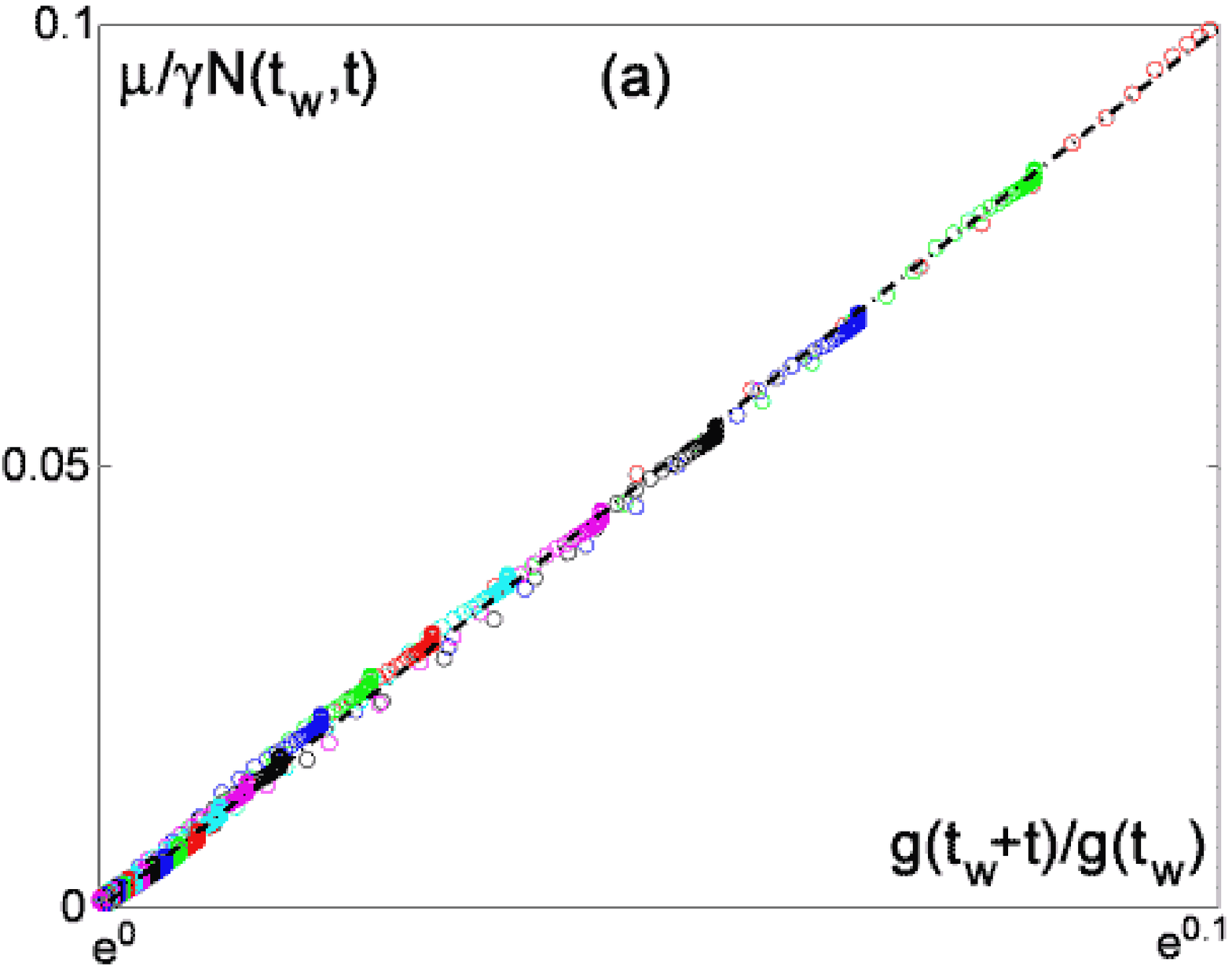}
\includegraphics[width=8cm]{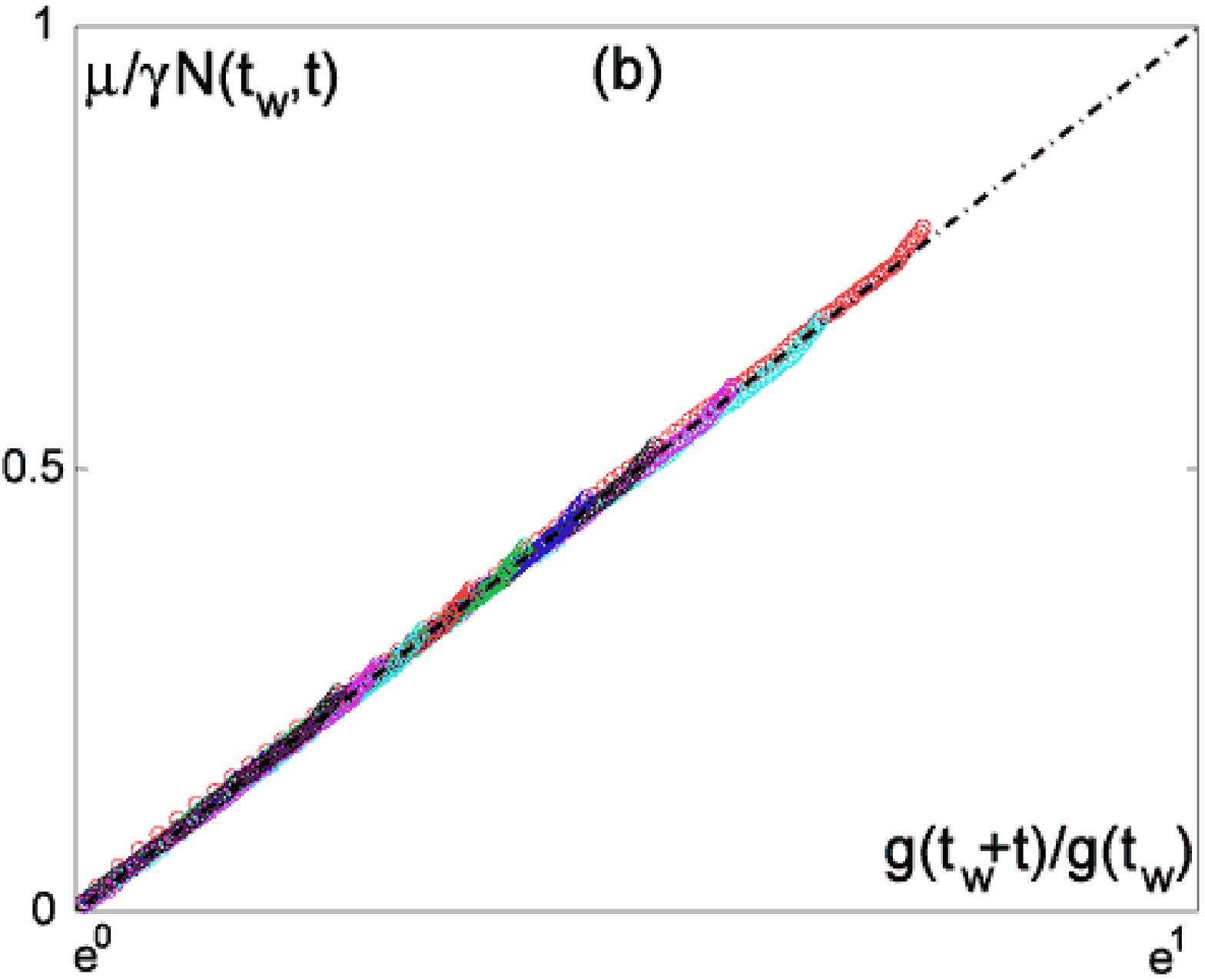}
\caption{$\frac{\mu}{\gamma}N(t_w,t)$ as a function of $g(t_w+t)/g(t_w)$ for each type of relaxation: (a) the exponential decay and (b) the intermittent dynamics}
\label{Nscaled}
\end{figure}

\noindent
We further estimate $P_m(0)$ by imposing the ratio $\eta=\frac{\gamma}{\alpha}$ in the case of the simple exponential decay. This ratio is nothing but the averaged number of move per bead and can reasonably be set to $\eta=1$ by considering that the exponential decay dynamics is precisely obtained when each bead simply relaxes once from the active state to the inactive one. The obtained value $P_m(0)=0.54$ is then imposed {\it a priori} to both dynamics. This allows us to determine the elementary timescales $1/\alpha$ and $1/\delta$ for both dynamics as well as $\eta$ in the intermittent dynamics (see right of Table~\ref{paramtab}). These obtained values are in very good agreement with the direct observations: $1/\alpha$ varies typically between $50 {\rm s}$ and $300 {\rm s}$ as given by the direct measurement of the exponential decay rate displayed on fig~\ref{times}(a). In the same way, $1/\delta$ is perfectly representative of the total duration of the relaxation process, namely of the order of $1/\alpha$ in the case of the exponential decay and larger than $10^3 s$ when intermittent bursts occur. Finally, $\eta=4.7$ is a reasonable value for the average number of move per bead in the sense that it agrees well with the average number of burst during an intermittent relaxation. Altogether, despite its simplicity, our model based on elementary dynamical processes provides a good description of the relaxation dynamics.

\section{Discussion}

Let us come back to the model properties and comment the analytical expression of the correlation function $C(t_w,t)$, that we obtained. As mentioned above, this correlation is the sum of two contributions, a first one which decays exponentially on a characteristic time $1/\gamma$, and a second one with a more complex form, which decays a priori on a longer time scale which depends on $t_w$. Thus, one could expect $C(t_w,t)$ to exhibit the typical two-step relaxation form familiar to the glass community. However, the first step of the relaxation is hard to evidence since its relative contribution is given by $P(t_w)$ which vanishes as $t_w$ increases, so that we mainly focus on the second step of the relaxation. The corresponding term presents a particular scaling form, as it depends only on the rescaled function $g(t_w+t)/g(t_w)$ -- this latter ratio also appearing in the expression of $N(t_w,t)$. Interestingly, this scaling form is precisely the generic aging form proposed by Cugliandolo and Kurchan \cite{Cugliandolo94}, which generalizes the simple aging scaling in $(t_w+t)/t_w$ by introducing a reparametrization of time $g(t)$. A natural question in this context is to identify the characteristic time scale associated to the decay of $h(t_w,t)=[g(t_w+t)/g(t_w)]^{-1}$, for a given $t_w$. This characteristic time $t_0$ can be identified with the inflexion point of $h(t_w,t)$ when plotted as a function of $\ln t$. Figure~\ref{ageing} displays the relaxation of this aging-like term, for several $t_w$ and for two different sets of parameters. 

Several regimes have to be distinguished according to the value of $t_w$. For $t_w \ll 1/\delta$, one finds $t_0 = 1/\mu P_0 + t_w$ (which can be approximated by $1/\alpha + t_w$), whereas for $t_w \gg 1/\delta$, $t_0$ saturates to the value $1/\delta$.
On figure~\ref{ageing}(a), the parameters are set to arbitrary values such that time scales are widely separated, so as to evidence the three regimes described above, namely $t_0 \approx 1/\alpha$, $t_0 \approx t_w$ and $t_0 \approx 1/\delta$. One clearly observes the simple
aging behaviors (i.e. $t_0 \approx t_w$) in the time window $1/\alpha \ll t_w \ll 1/\delta$, the earlier and later time regimes being stationary. One may also notice the strong
increase of the plateau value when $t_w$ reaches $1/\delta$  due to the fact that $g(t)$ \emph{saturates} for large $t$ instead of diverging. The scenario in the present model thus appears  a bit different from the full aging case.

\begin{figure}[htb]
\includegraphics[width=8cm]{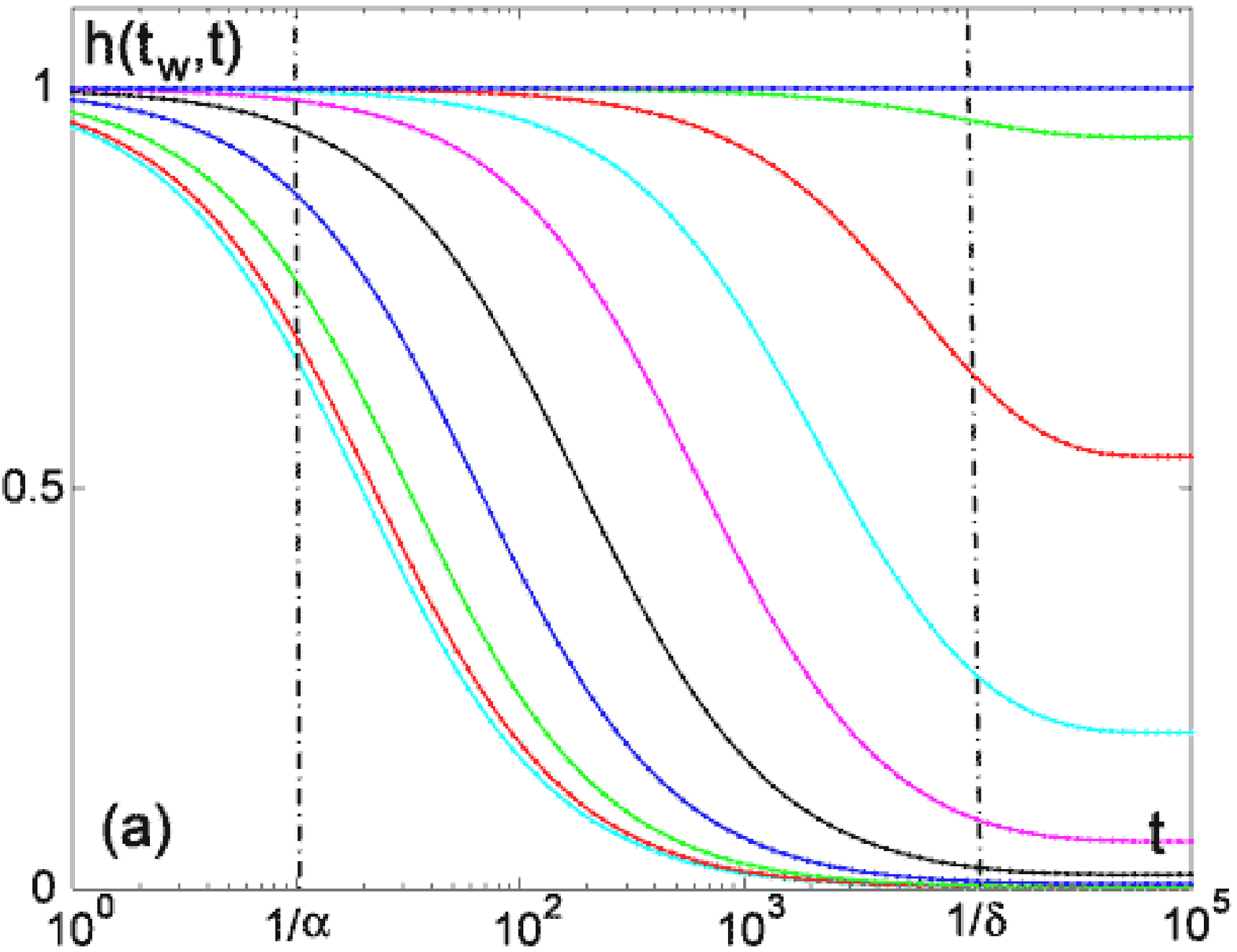}
\includegraphics[width=8cm]{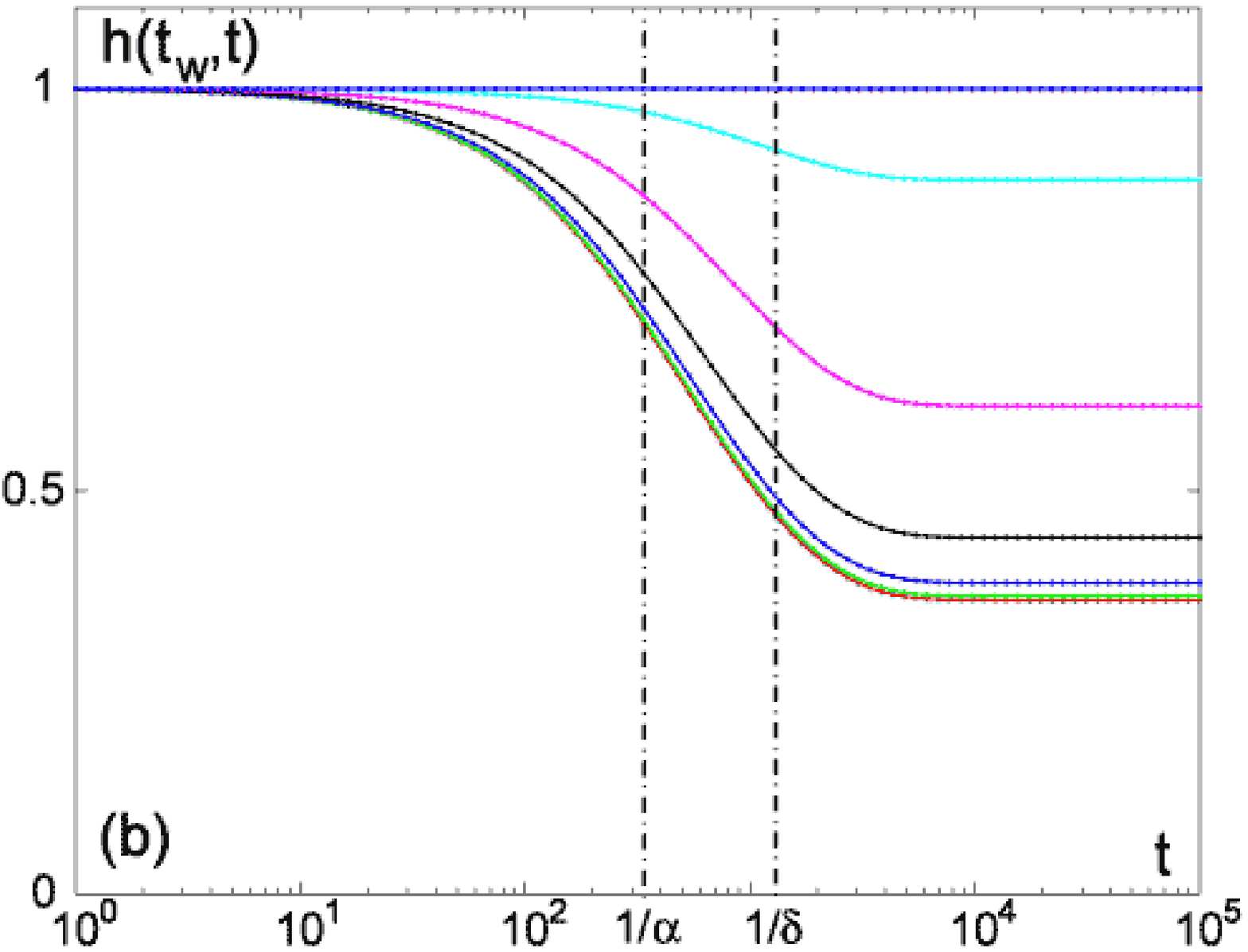}
\caption{$h(t_w,t)$ as a function of $\ln t$, for different $t_w$ for two set of parameter values: (a) $1/\alpha=10 {\rm s}$, $1/\delta=10^4 {\rm s}$ and (b) $1/\alpha=294 {\rm s}$, $1/\delta=1250 {\rm s}$.}
\label{ageing}
\end{figure}

\noindent
Figure~\ref{ageing}(b) displays the same function using the parameters values given in table~\ref{paramtab}, in order to extrapolate to large times the experimental data. The results are rather similar, in particular concerning the behaviors of the plateau, even if the intermediate aging regime is much less obvious, because of the weaker time-scale separation between $1/\alpha$ and $1/\delta$.

This saturation has two important consequences: on one hand, aging must be interrupted beyond a certain time scale, and on the other hand the correlation function does not decay to zero, but rather to a plateau value depending on $t_w$, which goes to $1$ when $t_w \to \infty$. Although this behaviors might seem surprising at first sight, its interpretation is clear: the global activity of the system vanishes at long times, so that decorrelation is harder and harder to achieve. Aging is interrupted in the present case because of a complete dynamical arrest, whereas in (thermal) glassy systems aging ceases if the system reaches Boltzmann equilibrium, thus allowing for further decorrelation. One can expect this behaviors to be typical of athermal jamming systems which are not driven by an external noise, but rather by a self-generated one, and for which the dynamics dies at long times. In this framework, the vanishing of the dynamics and the resulting interruption of aging is then encoded in the saturation of the time reparametrization $g(t)$. This may be considered as a general scenario for the long time dynamics of (non driven) athermal jamming systems, like spontaneously relaxing granular piles.

\section{Conclusion}

In this paper, we have studied the behaviors of a granular pile in a rotating drum after avalanches. We briefly consider the statistics of the angle of the pile $\theta(t)$ in the avalanching regime, measuring in particular the histograms of the starting angle $\theta_{start}$ and of the stopping angle $\theta_{stop}$. We also characterize the effect of a localized perturbation, showing that the perturbed area strongly increases with the angle $\theta$ at which the experiment is performed.

Our main results concern the long time mechanical relaxation of a granular pile just after a avalanche has occurred. Quantifying the `activity' in the pile $\delta A(t)$ by comparing subsequent images, we show that two different kind of relaxations are found, namely monotonous exponential decays with a characteristic time of order $10^2 s$ and intermittent relaxations for which spontaneous correlated moves, or activity bursts, generate long transients lasting for more than $10^3 s$. These long lived rearrangements tend to be localized in a subsurface layer of thickness around $10$ or $20$ beads diameters. Interestingly, whereas the typical time $\tau_b$ elapsed between two bursts remains constant when varying the slope $\theta$ of the pile, the exponential relaxation time $\tau_{\downarrow}$ depends strongly on $\theta$, with a divergence of the form $(\theta_r-\theta)^{-1}$ when $\theta$ approaches the angle of repose $\theta_r$. Both kinds of relaxations can be found at least for $\theta \gtrsim 5^{\circ}$, and the probability to find an intermittent decay increases with $\theta$.

We have then introduced a two-time quantity $N(t_w,t)$ measuring the average number of detected moves between times $t_w$ and $t_w+t$, which allows for a precise analysis of the intermittent relaxation process. Using the predictions of a simple stochastic model that we also introduce, with three fitting parameters, we find a very good rescaling of $N(t_w,t)$ for all values of $t_w$, keeping the same set of parameter values. Theses values of the fitting parameters compare very well with what can be extracted directly from $\delta A(t)$. Moreover, computing a natural correlation function in the model in order to make contact with more familiar two-time quantity, we show that this correlation behaves essentially in an aging-like manner, as a function of $g(t_w+t)/g(t_w)$, where $g(t)$ can be interpreted as a reparametrization of time, which saturates for large times so as to account for the full dynamical arrest.

\bibliographystyle{unsrt}
\bibliography{bibGlass,bibGranul}

\end {document}